\documentclass[twocolumn]{aastex61}

\newcommand{\gaia}{{\it Gaia}}
\newcommand{\cosmic}{{\tt{COSMIC}}}
\newcommand{\bse}{{\tt{BSE}}}
\newcommand{\kms}{{\rm{km/s}}}
\newcommand{\perYr}{{\rm{yr^{-1}}}}
\newcommand{\yr}{{\rm{yr}}}
\newcommand{\Gyr}{{\rm{Gyr}}}
\newcommand{\kpc}{{\rm{kpc}}}
\newcommand{\pc}{{\rm{pc}}}
\newcommand{\thin}{{\rm{thin}}}
\newcommand{\thick}{{\rm{thick}}}
\newcommand{\BH}{{\rm{BH}}}
\newcommand{\LC}{{\rm{LC}}}
\newcommand{\BHLC}{{\rm{BH-LC}}}
\newcommand{\au}{{\rm{au}}}

\newcommand{\orb}{{\rm{orb}}}
\newcommand{\obs}{{\rm{obs}}}

\newcommand{\edits}[1]{\textcolor{black}{#1}}
\shorttitle{Revealing black holes with \textit{Gaia}}
\shortauthors{Breivik, Chatterjee, and Larson}

\begin{document}

\title{Revealing black holes with \textit{Gaia}}

\email{katelyn.breivik@northwestern.edu, sourav.chatterjee@northwestern.edu}

\author[0000-0001-5228-6598]{Katelyn Breivik}
\affil{ Department of Physics \& Astronomy, Northwestern University, Evanston, IL 60202, USA}
\affil{ Center for Interdisciplinary Exploration \& Research in Astrophysics (CIERA), Evanston, IL 60202, USA}
\author[0000-0002-3680-2684]{Sourav Chatterjee}
\affil{ Department of Physics \& Astronomy, Northwestern University, Evanston, IL 60202, USA}
\affil{ Center for Interdisciplinary Exploration \& Research in Astrophysics (CIERA), Evanston, IL 60202, USA}
\author{Shane L. Larson}
\affil{ Department of Physics \& Astronomy, Northwestern University, Evanston, IL 60202, USA}
\affil{ Center for Interdisciplinary Exploration \& Research in Astrophysics (CIERA), Evanston, IL 60202, USA}
\nocollaboration



\begin{abstract}

We estimate the population of black holes with luminous stellar companions (BH-LCs) in the Milky Way (MW) observable by \gaia. We evolve a realistic distribution of BH-LC progenitors from zero-age to the current epoch taking into account relevant physics, including binary stellar evolution, BH-formation physics, and star formation rate, to estimate the BH-LC population in the MW today. 
We predict that \gaia\ will discover between $3,800$ and $12,000$ BH-LCs by the end of its $5\,\yr$ mission, depending on BH natal kick strength and observability constraints. We find that the overall yield, and distributions of eccentricities and masses of observed BH-LCs can provide important constraints on the strength of BH natal kicks. \gaia-detected BH-LCs are expected to have very different orbital properties compared to those detectable via radio, X-ray, or gravitational wave observations. 

\end{abstract}


\section{Introduction} \label{sec:intro}
Recent discoveries of merging binary black holes (BBH) by the LIGO-Virgo collaboration have reignited the interest in the formation of stellar-mass BHs. Stellar-mass BHs are naturally created as the final state of stars with masses in excess of $\sim10M_{\odot}$. However, discovering these dark remnants remains notoriously hard. Currently employed channels for discovery of BHs require selective binary architectures. For example, BHs can be detected while accreting at a high enough rate from a close binary companion via X-ray and radio observations \citep[e.g.,][]{McClintock2006,Remillard2006}. Otherwise, BBHs, in even tighter orbits may emit gravitational wave (GW) radiation and be observed by GW detectors. 

Both methods introduce selection biases and leave a vast regime in parameter space of binaries involving at least one BH completely untapped. \edits{Based on the \citet{Kroupa2001} initial stellar mass function (IMF) and stellar mass in the Galaxy, the MW is expected to contain $\sim10^5$ BHs. Depending on the initial binary fraction of high-mass stars, a large fraction of these BHs are expected to have luminous companions \citep{Sana2012}.} Nevertheless, the BlackCAT catalog contains only $59$ BH candidates in the MW (and $5$ extragalactic) discovered via X-ray and radio observations \citep{Corral-Santana2016}. 

More recently the LIGO-VIRGO observatories have opened a new window to observe BHs by detecting GWs from merging BBHs \citep{GW150914,GW151226,GW170104,GW170814}. Estimated theoretical event rates based on $4$ detections and one lower-significance trigger event to date indicate that this channel will yield $\sim300-800$ by the end of the nominal \gaia\ mission \citep{Abbott2016LRR,GW170104}. However, these events are likely to be dominated by higher-mass and distant BBHs due to selection bias \citep[e.g.,][]{Mapelli2017}. Stellar-mass BHs are also expected to be observable with mHz GW frequencies using LISA \citep{Amaro2017}. The estimated number of BBHs in the MW observable by LISA is $\sim$several dozens \citep[e.g.][]{Christian2017,Sesana2016,Belczynski2010b,Nelemans2001}. 

\gaia\ provides a new opportunity to hunt for BHs with luminous stellar companions (LC) in the MW with orbital properties in a different regime relative to the methods discussed above. We emphasize that the required observations demand no additional resource allocation and come as a `bonus' to the science goals of \gaia.

The method of detection, in principle, is very simple. By mission end, \gaia\ will constrain the motions of $\sim10^9$ luminous stars in the MW \citep{Gaia2016b}. If some of these stars are in close binary orbits with BHs (orbital period comparable to the \gaia\ mission length), then the period of the binary and the mass of the BH can be estimated from the stellar motion. As a `proof of concept,' \citet{Mashian2017} recently suggested that \gaia\ may be able to detect $\sim\,f\times10^5$ such BHs over the nominal mission lifetime. Here, $f$ contains {\em several} important astrophysical considerations including distribution of natal kicks to BHs, binary fraction in massive stars, mass-transfer events, and the orientation of these binaries in the MW with respect to us. These astrophysical considerations are crucial in determining the number and properties of BH-LCs detectable by \gaia. 

In this work, we use a detailed population synthesis, taking into account astrophysical processes relevant for the formation of BH-LCs, to predict the number and properties of those \gaia\ will discover. We pay particular attention to how different assumptions for largely uncertain BH-formation physics may affect these results. In \S\ref{Sims}, we describe the setup of our simulations and detail observability cuts applied to our results to obtain the \gaia-observable population. We show key results from our simulations 
in \S\ref{Results}. We conclude and discuss implications of our results in \S\ref{Discussion}.

\section{Simulations}\label{Sims}
For this study, we consider LCs at any point during their main sequence (MS) or post-MS evolution, as long as they are luminous at the present epoch. We simulate populations of binaries with at least one massive star, the BH progenitor, from zero-age to the present epoch of the MW and collect the BH-LC binaries at present. We model both the thin and thick disks in the MW using our binary population synthesis code, \cosmic, which uses 
\bse\ to evolve binaries \citep{Hurley2002}. We have updated \bse\ to incorporate our latest understanding of stellar winds from high-mass stars \citep{Vink2001,Vink2005, Belczynski2010}, and natal kicks to BHs \citep{Fryer2012}. We also include core-mass dependent stellar envelope binding energies in our common envelope evolution prescription \citep{Xu2010}. See \citet{Rodriguez2016a} for a detailed description of all improvements. 


We initialize binary populations for the MW thin and thick disks using standard assumptions. 
We assume a \citet{Kroupa2001} IMF between $0.1$ and $100\,M_{\odot}$ for primary masses in both the thin and thick disks. Mass ratios are sampled uniformly between 0.001 and 1 \citep{Mazeh1992,Goldberg1994}. We assume a primary-mass-dependent binary fraction proportional to $1/2+1/4(\log\,m)$ \citep{Haaften2013}. We distribute the orbital separations uniformly in log-space up to $5.75\times10^{6}R_{\odot}$ and as $(a/10_{\odot})^{1.2}$ below $10\,R_{\odot}$ \citep{Han1998}. We assume a thermal eccentricity distribution \citep{Heggie1975}.

We assume a constant star formation rate (SFR) of $2.15\,M_{\odot}\perYr$ over $10\,\Gyr$ for the thin disk producing a total mass of $M_{\thin} = 2.15\times10^{10}\,M_{\odot}$ and a single burst of star formation $11\,\Gyr$ ago for the thick disk producing a total mass of $M_{\thick} = 2.6\times10^{9}\ M_{\odot}$ \citep{Robin2003}. \edits{We assign metallicity 
$Z=Z_{\odot}$ and $0.15Z_{\odot}$ for the thin and thick disks, respectively \citep{Yoshii2013}.}

\subsection{Natal Kicks to BHs}
\label{S:kicks}
The strength of BH natal kicks is uncertain, but is expected to be related to neutron star (NS) natal kicks. Detailed modeling of Galactic scale heights of individual observed BH X-ray binaries (XRB) suggest wide ranges in kick magnitudes. Similar modeling for Galactic scale heights of observed BH XRBs as a population indicates natal kicks of $\sim100\,\kms$ \citep[and references therin]{Repetto2017}. Depending on the orientation and magnitude of the natal kick, the orbital eccentricity may increase, decrease, or the binary may become unbound altogether. Since the detectability of a BH-LC using \gaia\ crucially depends on the sky-projected size of its orbit, BH natal kicks are expected to affect \gaia's detectability of BH-LCs. 

To this end, we simulate three sets of models using BH natal kick prescriptions bracketing possibilities in nature: zero natal kicks (Zero-kick), NS kicks modulated by fallback of mass onto the BH (FB-kick), and full NS kicks (NS-kick). \edits{Natal kicks are randomly oriented and NS natal kicks are drawn from a Maxwellian distribution with $\sigma = 265\,\kms$ \citep{Hobbs2005}.}

We break the simulation into three components: \textit{fixed population}, \textit{Galactic realizations}, and \textit{observational cuts}. The fixed population describes the distributions of binary properties resulting from population synthesis adopting a specific binary evolution model and star formation history. Galactic realizations are created by sampling from the fixed population, where each binary sampled is also assigned a Galactic position and orientation. Finally, observational cuts are placed on the Galactic population to determine which BH-LCs are detectable by \gaia. We detail each step below. 

\subsection{Fixed population}\label{subsec:fixed}
The fixed population encapsulates the full range of binary parameters resulting from a population synthesis using a given binary evolution model, BH formation physics, and star formation history; e.g., we generate a fixed population for the thin disk evolved with the FB-kick model. \edits{Our full suite of simulations results in $6$ fixed populations: three kick models each for the thin and thick disk populations.}

To ensure convergence is achieved even for the low probability regions of the parameter space, we employ a \textit{match} criteria that compares the normalized histograms of binary parameters between consecutively and cumulatively simulated populations. 
The \textit{match} is computed bin by bin and summed over the entire parameter range as:
\begin{equation} \label{eq: match}
 match = \frac{\sum\limits_{k=1}^{N} P_{k,i} P_{k,i+1}}{ \sqrt{ \sum\limits_{k=1}^{N} (P_{k,i}P_{k,i})\sum\limits_{k=1}^{N} (P_{k,i+1}P_{k,i+1})}},
\end{equation}
\noindent where $P_{k,i}$ denotes the probability for the $k^{th}$ bin for the $i^{th}$ iteration. 

As the number of simulated binaries increases, the \textit{match} tends to unity. We continue to run simulations until $match>0.999$. For this study, we compute the \textit{match} for the binary masses ($M_{\BH}, M_{\LC}$), orbital period ($P_{\orb}$), eccentricity ($\rm{Ecc}$), LC temperature ($T_{\LC}$), and LC luminosity ($L_{\LC}$). 

While generating the fixed population, we log the total sampled mass, including single stars, as well as birth and death rates of BH-LC binaries. From these rates, we compute the total number of BH-LC binaries, $N_{\BHLC}$, at present in each Galactic component by normalizing the total sampled mass of our fixed population to the total mass of the Galactic component.

\subsection{Galactic Realizations}\label{subsec:gxReal}
We generate a six-dimensional kernel density estimate (KDE) from the fixed population using parameters of interest ($M_{\BH}, M_{\LC}$, $P_{orb}$, $Ecc$, $T_{\LC}$, $L_{\LC}$). 
We sample binaries from this KDE and assign Galactocentric three-dimensional positions ($x_{Gx}, y_{Gx}, z_{Gx}$). The positions are drawn from simple spatial distribution functions representing the distribution of stars in the MW \citep{Yu2015}.  
For the thin disk, we assume all binaries are distributed with an exponential radial ($R$) fall-off and a hyperbolic secant dependence in the vertical direction ($z$):
\begin{equation}
\rho(R,z)\propto\exp^{-R/R_{\thin}}sech^2(-z/z_{\thin}),
\end{equation}
where $R_{\thin}=2.5\,\kpc$ and $z_{\thin}=0.352\,\kpc$. For the thick disk we assume an exponential radial and vertical fall-off:
\begin{equation}
\rho(R,z)\propto\exp^{-R/R_{\thick}}\exp^{-z/z_{\thick}},
\end{equation}
with $R_{\thick}=2.5\,\kpc$ and $z_{\thick}=1.158\,\kpc$. In both cases, we assume azimuthal symmetry with $\phi$ sampled uniformly between $[0,2\pi]$. We compute the heliocentric distance to each binary by assuming a solar position of $R=8.5\,\kpc$, $\phi=0.0$, and $z=20\,\pc$ \citep{Yoshii2013}. 

We randomly orient each binary with respect to a fixed observer by sampling inclinations ($i$) uniformly in $\cos(i)$ between $[-1,1]$, and argument of periapsis ($\omega$) and longitude of ascending node ($\Omega$) uniformly between $[0,2\pi]$.

We repeat the process detailed above to generate 500 Galactic realizations per model/star formation to explore variance in the \gaia-observable BH-LC population.

\subsection{Observability Cuts}\label{Cuts}

\begin{figure}
\plotone{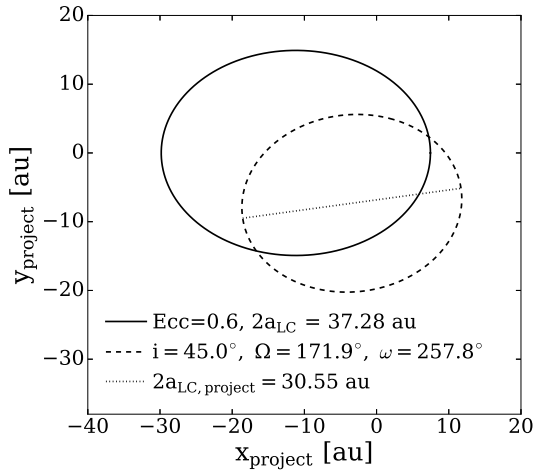}
\caption{\label{fig:orbit} Orbit projection for an example BH-LC with $M_{\BH}=10\,M_{\odot}$, $M_{\LC}=5\,M_{\odot}$, $a_{\LC}=37.28\,\au$, and $Ecc=0.6$. The solid black line shows the original orbit, with no prescribed orientation. The dashed line shows the projected size of the same orbit with binary orientation given by $i=45\,\deg$, $\Omega=171.9\,\deg$, and $\omega=257.8\,\deg$. The dotted line shows the projected semimajor axis $a_{\LC,\rm{project}}$.}
\end{figure}

We considered limiting magnitudes of $G=12-20$ based on expected \gaia\ performance. 
We compute visual magnitudes, $m_{\rm{V}}$, and $B-V$ colors from the bolometric corrections in \citet{Flower1996,Torres2010} using $L_{\LC}$ and $T_{\LC}$. We compute \gaia\ $G$ magnitude using the color-color transformations of \citet{Jordi2010}. 

We conservatively require observation of a full orbit during the \gaia\ mission lifetime, 
thus placing a hard upper limit $P_{orb}\leq5\,\yr$.  
We place a lower limit on the astrometric signature 
\begin{equation}\label{eq:alpha}
\alpha=\Big(\frac{a_{{\LC},\rm{project}}}{\au}\Big)\Big(\frac{d}{\pc}\Big)^{-1}\,\rm{arcsec},
\end{equation}
based on the projected size of the LC orbit ($a_{{\LC},\rm{project}}$),
such that 
$\alpha$ is greater than \gaia's magnitude-dependent astrometric precision, $\sigma_{G}(G)$ 
\citep{Gaia2016a}. 
We estimate optimistic (pessimistic) yields 
using $\alpha\geq\sigma_{G}$ ($3\sigma_{G}$). 
%
Binary separation is defined as $a=a_{\LC}+a_{\BH}=a_{\LC}(1+M_{\LC}/M_{\BH})$. Projected separation is computed with orbital parameters: $i$, $\Omega$, and $\omega$ using  the Thiele-Innes elements:

\begin{eqnarray} \label{eq:TIelements} 
 A &= a_{\LC}(\cos\  \omega\  \cos\  \Omega -  \sin\  \omega\  \sin\  \Omega\  \cos\ i)\\
B &=  a_{\LC}(\cos\  \omega\  \sin\  \Omega\ +  \sin\  \omega\  \cos\  \Omega\  \cos\ i)\\
F &=  a_{\LC}(-\sin\ \omega\ \cos\ \Omega\ -  \cos\ \omega\ \sin\ \Omega\ \cos\ i)\\
G &=  a_{\LC}(-\sin\  \omega\  \sin\ \Omega\ +  \cos\ \omega\ \cos\ \Omega\ \cos\ i).
\end{eqnarray}
The projected cartesian components $x_{\rm{project}}$ and $y_{\rm{project}}$ are:
\begin{eqnarray}
x_{\rm{project}} &= AX+FY \label{eq: xTrans}\\
y_{\rm{project}} &= BX+GY \label{eq: yTrans}, 
\end{eqnarray}
where $X=\cos\,E-Ecc$ and $Y =\sqrt{1-Ecc^2}\sin\,E$. The projected orbital separation, $a_{\LC,\rm{project}}$ is the semimajor axis of the projected ellipse. %

Figure\ \ref{fig:orbit} shows an illustration of these projection effects for a sample BH-LC. In this example, $M_{\BH}=10\,M_{\odot}$, $M_{\LC}=5\,M_{\odot}$, eccentricity $Ecc=0.6$, $P_{orb}=4\,\yr$, and, $a_{\LC}=18.64\,\au$. The projected separation, after transforming as in Eqs. \ref{eq: xTrans} and \ref{eq: yTrans} is 
$a_{\LC,\rm{project}}=15.28\,\au$. 

\section{Results}\label{Results}
\subsection{Full BH-LC Population}\label{NoCuts}
\begin{figure*}
\plotone{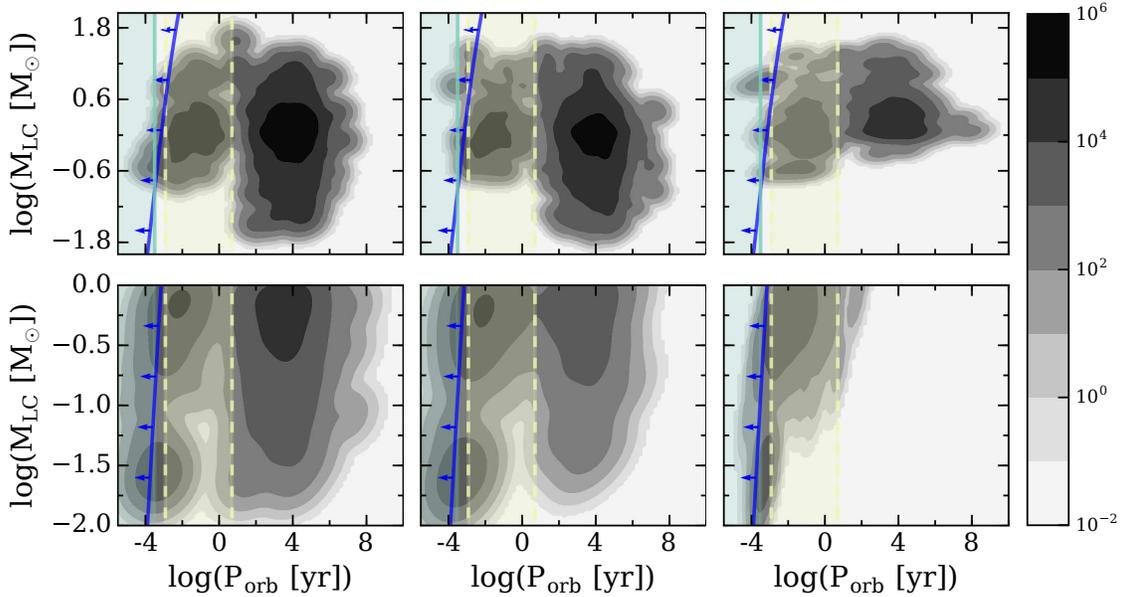}
\caption{\label{fig:porbVsMbhms} Two-dimensional PDFs for all BH-LCs in the MW in the LC-mass vs orbital period plane.  
Top and bottom rows show distributions for the thin and thick disk, respectively. Columns (a), (b), and (c) show Zero-kick, FB-kick, and NS-kick models, respectively. Light blue bands show the spread of orbital periods observable by \textit{LISA}. Blue lines represent an approximate upper limit for accreting BHs via Roche-lobe overflow (estimated using $M_{\BH}\approx36\,M_{\odot}$, the most massive BH created in our models). 
Yellow bands bounded by dotted lines show the \gaia-observable range, spanning $P_{\orb}\simeq0.5\,\rm{day}$ to $5\,\yr$.}
\end{figure*}

We consider first the overall population of BH-LCs at present, without observability cuts. This population represents the full range of BH-LC binaries potentially produced in the MW, for each of our models. Figure\ \ref{fig:porbVsMbhms} shows distributions of LC mass and orbital period of all BH-LC binaries from our simulations for BH-LCs containing both MS and post-MS stars. Light blue bands and blue lines show the general regions where BH-LCs may be observable by \textit{LISA} and X-ray observations respectively, while yellow dashed lines and bands show the region of orbital periods roughly observable by \gaia, with an illustrative lower limit placed at $P_{orb}\simeq0.5\,\rm{day}$. The lower limit represents the average minimum orbital period observable by \gaia\ from our optimistic models. 

The majority of BH-LCs with $P_{\orb}\leq5\,\yr$ lie in the \gaia-observable region. Only $\approx1\%$ BH-LC systems in the thin disk with $P_{orb}\leq5\,\yr$ also have $P_{orb}<0.5\,\rm{day}$, 
while $\approx11\%$ of thick-disk BH-LCs with $P_{\orb}\leq5\,\yr$ have $P_{\orb}<0.5\,\rm{day}$. Thus \gaia\ will provide an important probe into the BH-LC parameter space that is complementary and otherwise inaccessible to radio, X-ray and GW observations.

Note that BH-LCs with $P_{\orb}\gtrsim10^5\,\yr$ would likely be disrupted due to Galactic tides and appear in our full population (Figure\ \ref{fig:porbVsMbhms}) only because we did not take this into account. This does not affect our results for the \gaia-detectable BH-LCs since those are tightly bound and are unlikely to be disrupted unless placed in very dense environments near the Galactic center. 

The number of BH-LCs decreases with increasing natal kick magnitudes since larger kicks disrupt a larger fraction of primordial binaries. For the same reason, natal kicks also affect the distribution of $P_{\orb}$ of BH-LCs.  
While BH-LCs are disrupted in the Zero-kick model only if the change in mass during BH formation is severe enough to reduce the orbital binding energy to the point of unbinding, the NS-kick model leads to disruption of BH-LC progenitors with $P_{\orb}\gtrsim5\,\yr$. 


We find a significant population of BH-LCs with $P_{\orb}$ longer than the nominal \gaia\ mission in all models except the NS-kick model. \edits{This population exclusively contains BH-LCs that have not experienced a common envelope evolution. The population with orbital periods in the \gaia-observable range and lower contain BH-LCs that have experienced a common envelope, as well as BH-LCs that haven't.} Our NS-kick model does not allow formation of BH-LCs with $P_{\orb}\geq5\,\yr$ and $M_{\LC}\leq0.25\,M_\odot$, while other models with lower natal kicks allow formation of these systems (Figure\ \ref{fig:porbVsMbhms}). By extending the \gaia\ mission beyond the nominal $5\yr$, our $P_{\orb}\leq5\,\yr$ condition could be relaxed, and \gaia\ may constrain highly uncertain BH natal kicks from the properties of detected BH-LCs at longer orbital periods. 

\subsection{Gaia-Observable Population}\label{PopCuts}

\begin{figure}[t!]
\plotone{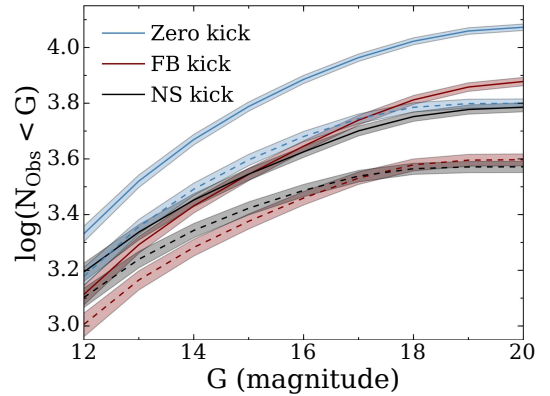}
\caption{\label{fig:NlimMag} Mean number of observable BH-LCs as a function of limiting $G$ magnitude computed from $500$ Galactic realizations. Shades show $3\sigma$ regions above and below the mean denoted by the lines. Dashed (solid) lines denote our estimates using pessimistic (optimistic) astrometric cuts (\S\ref{Cuts}). Blue, red, and black denote Zero-kick, FB-kick, and NS-kick models, respectively. 
}
\end{figure}
Figure\ \ref{fig:NlimMag} shows the number of systems observable by \gaia\ ($N_{\obs}$) as a function of limiting $G$ magnitude. The number of observable systems is inversely related to the BH natal kick strength because the likelihood to unbind a BH-LC progenitor increases with kick strength. 

\begin{table}
\caption{Average number of BH-LCs observable by \gaia\ from 500 Galactic realizations with $G<20$ for our optimistic and pessimistic cuts (\S\ref{Cuts}).
\label{tbl:AstroCut}}
\begin{center}
\begin{tabular}{ccccc}
\tableline
Model & Component & Total & Obs Opt & Obs Pess \\
\tableline
Zero-kick & Thin Disk & 475840 & 9120 & 5145\\
 & Thick Disk & 28197 & 2708 & 1159 \\
\tableline
FB-kick & Thin Disk & 250707 & 4632 & 2786\\
 & Thick Disk & 12523 & 2918 & 1170\\
\tableline
NS-kick & Thin Disk & 71653 & 4823 & 3249\\
 & Thick Disk & 6561 & 1283 & 480\\
\tableline
\tableline
 \end{tabular}
 \end{center}
\end{table}

Table\ \ref{tbl:AstroCut} shows the result of our astrometric cuts. The total number of BH-LCs {\em formed} in our models is roughly consistent with the estimates in \citet{Mashian2017}. However, the majority of the systems formed lie above our upper $P_{\orb}$ cut (Figure\ \ref{fig:porbVsMbhms}). In addition to the natal kick prescription, the Galactic component has a noticeable effect on the number of BH-LC systems. Since the thick disk is modeled as a single star burst $11\,\Gyr$ ago, all LCs in the thick disk at present are low-mass. Large natal kicks for BHs like our NS-kick model affect BH-LCs with low-mass LCs more (Figure\ \ref{fig:porbVsMbhms}; \S\ref{NoCuts}). As a result, in the thick disk, the NS-kick model yields a very low $N_{\obs}$. The adopted continuous SFR in the thin disk keeps $N_{\obs}$ relatively high even for the NS-kick model by forming high-mass luminous stars very close to the present epoch.  

We estimate $3,800-12,000$ BH-LCs in the MW are detectable by \gaia\ depending on adopted BH natal kick strengths and demand of the astrometric cut. Considering all cases, \textit{we predict at least $10^3$ BH-LCs observable by \gaia.}

Note that carefully considering astrophysics and \gaia's selection biases in our study has reduced the predicted yield by a factor of $\sim10^{-2}$ relative to the prediction by \citet{Mashian2017}. Still, our predicted yield is nearly two orders of magnitude higher than the number of BH binaries already known. Thus, \gaia\ observations could dramatically improve our understanding of BHs and BH binaries in particular. 

\begin{figure*}
\plotone{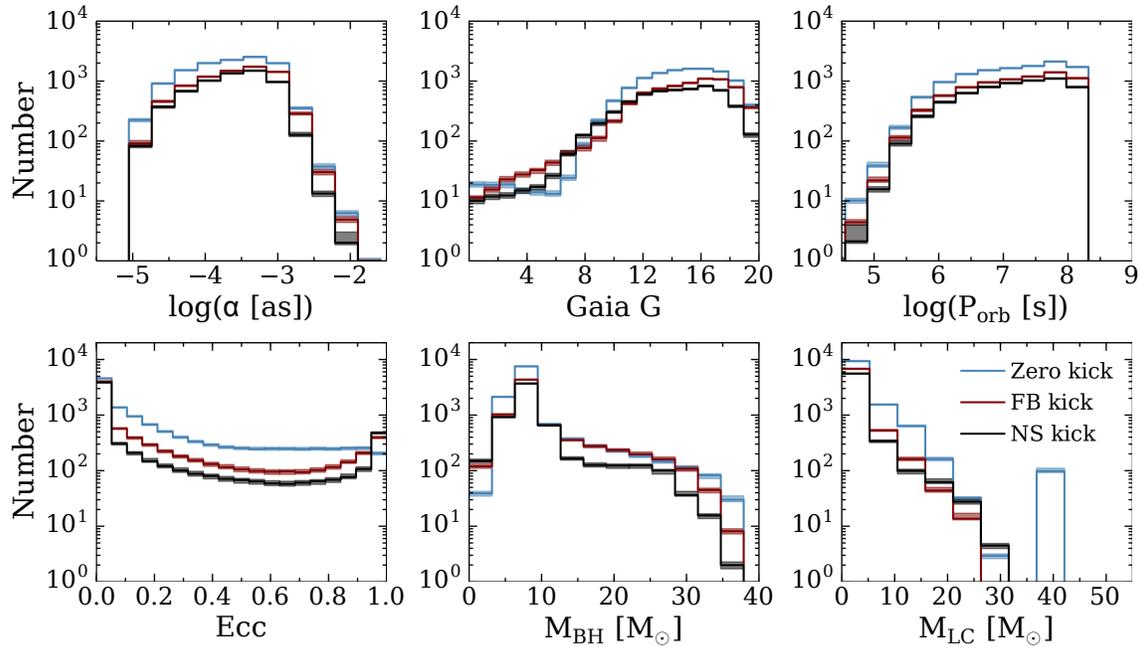}
\caption{\label{fig:ObsParams} Distributions for the number of observable properties of BH-LCs in the MW (thin and thick disks) potentially detectable by \gaia. Histograms represent the mode of each bin height for our $500$ population realizations, with shaded regions showing $3\sigma$ confidence limits above and below the mode. Blue, red, and black denote Zero-kick, FB-kick, and NS-kick models, respectively.} 
\end{figure*}

\subsection{Observable BH-LC Parameters}\label{Params}
Magnitude $G$, $\alpha$ (Eq. \ref{eq:alpha}), and $P_{\orb}$ will be directly observed for a \gaia-detected BH-LC, while binary masses and $Ecc$ can be inferred via modeling. Figure\ \ref{fig:ObsParams} shows distributions of these properties for observable BH-LCs in our models.  

Distributions of $\alpha$, $G$, and $P_{\orb}$, are governed primarily by hard requirements on $\alpha$ and $P_{\orb}$ we impose for detectability, and our adopted primordial IMF and binary properties (\S\ref{Sims}). 
Distributions of $Ecc$, $M_{\rm{BH}}$, and $M_{\LC}$ exhibit differences depending on the adopted BH formation physics. 

Most notable is the rise in the number of systems with $Ecc>0.9$ for the FB-kick and NS-kick models, while the Zero-kick model flattens at high eccentricities. This is a direct outcome of eccentricity excitation by SN kicks to BHs. Independent of the BH-physics model, a high fraction of detectable BH-LCs have high eccentricities. Thus, effects of eccentricities on the sky-projected size of a binary (Figure\ \ref{fig:orbit}) must be taken into account to correctly interpret astrometric observations of BH-LCs.

Mass distributions also vary to some degree from model to model. The NS-kick model shows a dearth of BHs with masses between $15\,M_{\odot}$ and $25\,M_{\odot}$ relative to the Zero-kick and FB-kick models. In this range BHs receive zero or little natal kicks for the Zero-kick and FB-kick models, respectively, whereas the BHs in the NS-kick model receive large natal kicks \citep{Fryer2012}. 
The $M_{\LC}$ distribution is affected similarly between $5M_{\odot}$ and $20M_{\odot}$, where models with larger natal kicks show fewer observable BH-LCs as a result of binary disruptions during BH formation. \edits{We do not expect the peak at $M_{\LC}\simeq 40\,M_{\odot}$ in the Zero-kick model to be an observable feature in a future \gaia\ BH-LC catalog. It is casued by a single outlier system in the Zero kick fixed population that is exceptionally bright due to it's mass. This system propagates into our observable synthetic data set during the Galactic realization sample, and always deemed observable based on the bright LC  and orbital period near $5\,\yr$.}

Admittedly, deriving masses and eccentricities with the observation of only $\alpha$, $G$, and $P_{\orb}$ is difficult. 
However, \gaia\ will provide low resolution spectral energy distributions in red and blue for all observed LCs as well as radial velocities for LCs 
brighter than $G=17$ \citep{Jordi2010}. Combining these observations with astrometric information may allow parameterization of the binary, including minimum-mass measurements of the BH. Such modeling will be the topic of a future study.

\section{Discussion}\label{Discussion}
In this study we have predicted that \gaia\ will potentially observe $\sim10^3-10^4$ BH-LCs in the MW depending on BH natal kick physics and observational constraints. 
These observations could increase the number of known BHs by two orders of magnitude in less than $5\,\yr$. Moreover, the properties of the BH-LCs \gaia\ will detect occupy a different regime from those detectable via currently employed methods (Figure\ \ref{fig:porbVsMbhms}).  

We have used a detailed population synthesis to show the effects of BH natal kicks on the Galactic BH-LC population. The number of observed systems places the strongest constraint on the natal kicks BHs receive: The number of observable BH-LCs decreases with increasing natal kick strength (Figure\,\ref{fig:NlimMag}). Moreover, we find that distributions of eccentricity and masses of observable BH-LCs may also help constrain natal kick physics, especially if \gaia's mission length is extended (Figures\ \ref{fig:porbVsMbhms}, \ref{fig:ObsParams}). 

We predict far fewer observable BH-LC systems than the $2\times10^5$ predicted by \citet{Mashian2017}, suggesting that the fraction of BH-LC progenitors that survive to the present is low ($f\sim0.01$). In spite of this, \gaia\ promises to discover at least $10^3$ BH-LCs.

A large fraction of LCs in the observed BH-LC population will evolve into compact objects. Thus, connections may be drawn between the \gaia-observable population and the X-ray- or GW-observable populations. Thus, BH-LCs observable by \gaia\ provide a unique opportunity to probe the population of stellar-mass BHs.

\acknowledgments

The authors gratefully acknowledge the anonymous referee for help in clarifying key points. KB and SLL acknowledge support from NASA Grant NNX13AM10G. SC acknowledges support from NASA issued by Hubble Space Telescope Archival research grant HST-AR-14555.001-A. 

%


\software{astropy \citep{2013A&A...558A..33A},  
         BSE \citep{Hurley2013}
          }

\end{document}